\documentclass[twocolumn,secnumarabic,amssymb, nobibnotes,showkeys, aps, pre]{revtex4-1}

\usepackage[ansinew]{inputenc}
\usepackage[english]{babel}
\usepackage{amsmath}
\usepackage{graphicx} 

\usepackage{natbib}

\DeclareMathOperator{\erfc}{erfc}

\begin{document}

\title{Fixed-density boundary conditions in overdamped Langevin simulations \\
of diffusion in channels}

\author{L. Ram\'irez--Piscina}
\affiliation{Departament de F\'isica, Universitat Polit\`ecnica de Catalunya,
Avinguda Doctor Mara\~n\'on 44, 08028 Barcelona, Spain}
\date{\today}

\begin{abstract}
We consider the numerical integration of Langevin equations for particles in a channel, in the presence of boundary conditions fixing the concentration values at the ends. This kind of boundary condition appears for instance when considering the diffusion of ions in molecular channels, between the different concentrations at both sides of the cellular membrane. For this application the overdamped limit of Brownian motion (leading to a first order Langevin equation) is most convenient, but in previous works some difficulties associated with this limit were found for the implementation of the boundary conditions. We derive here an algorithm that, unlike previous attempts, does not require the simulation of particle reservoirs or the consideration of velocity variables or adjustable parameters. Simulations of Brownian particles in simple cases show that results agree perfectly with theory, both for the local concentration values and for the resulting particle flux in nonequilibrium situations. The algorithm is appropriate for the modeling of more complex ionic channels and, in general, for the treatment of analogous boundary conditions in other physical models using first order Langevin equations.  ***This version corrects missprints in two equations of the published paper***
\end{abstract}

%\pacs{07.05.Tp,87.10.Mn, 87.16.Vy}


\maketitle

%\newpage

\section{Introduction}

Ionic channels are protein structures at the nanometric scale that permit the crossing of specific ions through the cell membrane, leading to important physiological functions \cite{Hille}. 
Ion flow through the channel is driven by the concentration difference between the extracellular and intracellular media, and by the action of electrostatic forces due to the membrane potential. 
Experiments performed on single channels show the presence of strong fluctuations, both in the ionic flux and in the gating dynamics of the channel \cite{Hammond}.
For the modeling of this problem it is often convenient to consider the diffusive motion of ions inside the channel, in combination of some specific dynamics for the gates \cite{Hille,levitt1986interpretation}. In some semimicroscopic approaches this is achieved by means of stochastic differential equations, or Langevin equations, for ion positions and other variables \cite{cooper1985theory,levitt1999,LEE200279,ramirez1}.
Langevin equations are employed for modeling a large variety of systems in which fluctuations are relevant for their dynamics \cite{gardiner2004,ojalvo99}.
In such equations, the dynamic variable obeys a stochastic differential equation in which most microscopic degrees of freedom are substituted by stochastic terms or noises.

We consider here the Langevin dynamics of a population of Brownian particles, representing the individual ions, moving in a molecular channel between two particle reservoirs with fixed concentrations, corresponding to those of both sides of the membrane.
The low Reynolds number values associated with such small scales imply that inertia is negligible and one can safely take the overdamped limit of the Langevin equation, which means that the problem can be formulated as a set of first order stochastic differential equations for the particle positions. 
%We will consider the overdamped limit of the Langevin equation, in which inertia is negligible and the problem is formulated as a first order differential equation for the particle position.

Whereas other boundary conditions on Brownian particles are standard and somewhat trivial in simulations (e.g. absorbing or reflecting boundary conditions \cite{gardiner2004}), fixed concentration conditions have turned out to be more difficult to implement efficiently in the overdamped case \cite{barcilon93,singer04}. Previous attempts involved for instance the simulation of particles in the reservoirs \cite{corry2002reservoir,gomez-marin}, the consideration of random walks as discrete-time Brownian dynamics \cite{schumaker2002boundary},
the employing of velocity variables \cite{eisenberg95,nadler05}, %(note however that for low Reynolds the system is at the overdamped limit and inertia effect should be negligible), 
or some shooting method to empirically finding simulation parameters \cite{singer05}.
Here we make explicit the derivation and procedure of a satisfactory method for these boundary conditions, without such shortcomings, recently implemented in ionic channel simulations % in a semimicroscopic approach for ionic channel dynamics
\cite{ramirez1,ramirez2,ramirez-periodic}. For simplicity, we will illustrate the method by the simulation of very simple cases, consisting of non-interacting particles in one-dimensional static channels devoid of any biological complexity. The treatment of the boundary conditions would be applicable to more complex and realistic situations.

The outline of this article is the following. In the next section we state the problem and show some known theoretical results, which will be useful for interpreting simulations. In Sec.~\ref{sec-algorithm} we derive the algorithm for the boundary conditions with fixed concentration values. In Sec.~\ref{sec-results} we perform numerical simulations in order to check the method. In view of the obtained results we also discuss some assumptions that are implicit in this kind of boundary conditions. We end with some conclusions.

\section{Langevin equation with fixed concentrations boundary conditions}
\label{sec-langevin}

We consider the temporal evolution of an ensemble of independent Brownian particles, moving in the overdamped limit in an external potential $V(x)$ through a channel of length $L$. For simplicity we consider here the one dimensional dynamics in the x-direction of the particles. The extension of the method to a three-dimensional system in contact with a surface with fixed concentration is immediate. The Langevin equation for the position $x\in (0,L)$ of each of the particles is then
\begin{equation}
 \gamma {\dot x} = -\frac{d V(x)}{d x} + \xi(t),
\label{eq-langevin}
\end{equation}
where $\gamma$ is a friction coefficient and $\xi(t)$ is a gaussian white noise with zero mean and correlation given by 
\begin{equation}
\langle \xi (t) \xi(t') \rangle = 
2 \gamma k_B T \delta (t- t').
\label{eq-correl}
\end{equation}
Noises acting on different particles are independent (uncorrelated). These equations have been constructed verifying the fluctuation-dissipation theorem, and then lead to the correct equilibrium solutions. Here we will put the system in a nonequilibrium situation by considering the channel connecting two reservoirs with fixed values of particle concentration placed at both ends of the channel, and adding an external field acting on the particles.

Let $\rho_1$, $\rho_2$ be the particle concentration (i.e. number of particles per unit length in the channel) at $x=0, L$ respectively. In the case of a very narrow channel connecting two large volumes, the values of $\rho_1$, $\rho_2$ are related to the three-dimensional concentrations $c_i$ (particles per unit volume at the reservoirs) by $\rho_i = A \, c_i$, with $A$ being the transversal section of the channel. Note that for a very narrow channel and large and well mixed reservoirs the flow of ions will have a negligible effect on the values of the concentrations at the reservoirs, which permits us to consider constant values for the boundary conditions. 
In order to have a theoretical reference for the simulation results it is convenient to consider the Fokker-Planck equation for the concentration $\rho(x,t)$ \cite{gardiner2004}:
\begin{equation}
 \label{eq-fokker}
 \frac{\partial\rho(x,t)}{\partial t}
 = - \frac{\partial}{\partial x} f(x) \rho(x,t) 
 + D \frac{\partial^2}{\partial x^2} \rho(x,t),
\end{equation}
with the drift velocity $f(x)$ and the diffusion coefficient $D$ given by
\begin{eqnarray}
 f(x) = - \frac{1}{\gamma}\frac{d V}{d x},\;
 D = \frac{k_B T}{\gamma}
\end{eqnarray}

% \begin{equation}
%  \label{eq-fokker}
%  \frac{\partial\rho(x,t)}{\partial t}
%  = \frac{1}{\gamma}\frac{\partial}{\partial x} \frac{d V}{d x} \rho(x,t) 
%  + \frac{k_B T}{\gamma} \frac{\partial^2}{\partial x^2} \rho(x,t).
% \end{equation}

The steady solution of Eq.~(\ref{eq-fokker}) with these boundary conditions can be written as
\begin{equation}
 \label{eq-steady}
 \rho(x) = e^{-V(x)/k_BT}\left(\rho_1 e^{V(0)/k_BT}-J\frac{\gamma}{k_BT}\int_0^x e^{V(x')/k_BT}dx' \right),
\end{equation}
with $J$ being the steady current of particles, which is given by
\begin{equation}
 J=\frac{k_BT}{\gamma}
 \frac
 {\rho_1 e^{V(0)/k_BT}-\rho_2 e^{V(L)/k_BT}}
 {\int_0^L  e^{V(x')/k_BT}dx'}.
\end{equation}

In the particular case of particles moving under the action of a constant drift (for instance charges $q$ inside a planar capacitor with voltage $\phi$, which constitutes a model for a membrane channel), we can write $V(x)=q\phi(x-x_1)/L$. Then the current is
\begin{equation}
 J=-\frac{q\phi}{\gamma L}\frac{\rho_1-\rho_2 \exp({q\phi/k_BT})}{1- \exp({q\phi/k_BT})}.
 \label{eq-current}
\end{equation}
For large voltages the current is controlled by the rate at which particles can enter into the channel, which translates into a dependence on one of the boundary
concentrations, depending on the sign of the voltage. That is, for $q\phi/k_BT\gg 1$ we have $J\simeq -q \phi \rho_2 / \gamma L$, 
whereas  for $q\phi/k_BT\ll -1$ we have $J\simeq -q\phi\rho_1/\gamma L$.
Between these two limiting regimes the system presents a crossover in which the current is regulated both by the different concentrations at both boundaries and by the value of the voltage. When checking numerical results it will be appropriate to consider cases from each of these three regimes.

\section{Numerical algorithm}

\label{sec-algorithm}

\subsection{Time-discretized Langevin dynamics}

Let us consider a temporal step $\Delta t$. An explicit Euler algorithm for the evolution of the position variable of the Brownian particle of Eq.~(\ref{eq-langevin}) calculates the new $x(t+\Delta t)$ by using the values at the old position $x(t)$:
\begin{equation}
\label{eq-euler}
 x(t+\Delta t) = x(t) + f(x(t))\Delta t + \sqrt{2D\Delta t} \; \chi,
\end{equation}
where $\chi$ is a Gaussian random number of zero mean and unit variance. 
%We assume that all particles in the system evolves in a discretized time following this dynamics. 
For later use we can also write an equivalent formulation by means of the conditional probability density
\begin{equation}
\label{eq-propagator}
 p(x,x')=\frac{1}{\sqrt{4\pi D \Delta t}}\exp - \frac{(x-x'-f(x')\Delta t)^2}{4D\Delta t},
\end{equation}
where $p(x,x')dx$ is the probability of a particle being at a position in the interval $(x, x+dx)$ at the end of the time step (at time $t+\Delta t$) if it was at position $x'$ at the beginning (at time $t$). 

This last expression corresponds, for a constant drift ($x$-independent $f$), to the result that can be directly obtained from the diffusive motion of the original Langevin [Eq.~(\ref{eq-langevin})]. Therefore Eq.~(\ref{eq-propagator}) is exact for any $\Delta t$ in such constant drift case. 

\subsection{Boundary conditions}

The boundary conditions will be implemented in the following way. The particles in the system evolve according to Eq.~(\ref{eq-langevin}) until at the end of any time step any of the particles escape from the limits of the system. In this moment the particle disappears from the simulation and it is never re-used. At the same time there exists a certain probability for new particles entering into the system through each boundary at each time step.

These new particles, inserted with a given probability, start their dynamics at some initial position (which has also to be determined), and evolve together with the rest of the particles until at some time step they could also exit from the system.
The objective here is to calculate this insertion probability and to determine the distribution of initial positions.

\subsubsection{Rate of new particles}

Let us consider the left boundary at $x=0$. The other boundary at $x=L$ can be treated in an equivalent way.
We consider the semi-infinite domain $(-\infty,0)$ (the particle reservoir) where there is supposed to exist an infinite number of virtual particles with a homogeneous concentration (particles per unit length) $\rho_1$, all evolving with the dynamics of Eq.~(\ref{eq-langevin}) but with a constant value of the drift equal to that of the boundary, i.e. $f=f(0)$. Some of these virtual particles can enter into the system and have a position $x>0$ at the end of the time step $\Delta t$. Let $\rho_{in}(x)$ be the local density inside the system of these new particles that in the previous step were outside the system. It can be calculated by using Eq.~(\ref{eq-propagator}) as
\begin{equation}
\label{eq-rhoin}
 \rho_{in}(x)= \int_{-\infty}^0 dx' \rho_1 p(x,x') = 
 \frac{\rho_1}{2} \erfc\left(\frac{x-f\Delta t}{\sqrt{4 D \Delta t}}\right).
\end{equation}
The mean number of particles entering through this boundary during the time step $\Delta t$ can be calculated by integrating this density. After some manipulations, and by using known recurrence relations for the integrals of the error function [see for instance Eq. (7.2.5) in Ref. \cite{abramowitz}], we get
\begin{equation}
\label{eq-nmean}
 \langle N \rangle = \int_0^\infty dx \rho_{in}(x) = \rho_1\sqrt{D\Delta t}\;q(a),
\end{equation}
where $a$ and the function $q(x)$ are given by
\begin{eqnarray}
 \label{eq-a}
 a &=& -{f}{\sqrt{\frac{\Delta t}{4 D}}}
\\
 \label{eq-q}
 q(x) &=& -x \erfc(x) + \frac{1}{\sqrt\pi}\exp -x^2.
\end{eqnarray}

Since the particle arrivals are independent events, the number of particles that enter into the system at each time step should be chosen from a Poisson distribution with the mean number $\langle N \rangle$ calculated in Eq.~(\ref{eq-nmean}). This can be done by using standard techniques \cite{press07}. 
However, for small $\Delta t$ and not very large densities, one has $\langle N\rangle \ll 1$ and the generation of the Poisson random number can be avoided. Note that for small $a$ Eq.~(\ref{eq-nmean}) reduces to 
$\langle N \rangle \simeq \rho_1\sqrt{D \Delta t / \pi} +\frac{1}{2}\rho_1 f \Delta t$
so this condition impose two conditions on $\Delta t$, namely $\Delta t \ll (\rho_1^2 D)^{-1}$ and $\Delta t \ll (\rho_1 f) ^{-1}$.
In such case one can neglect the possibility of two o more particles appearing during the same time step through the same boundary and identify the obtained  mean value with the probability of appearing one single particle, i.e. $p^{(1)} \simeq \langle N \rangle$. The algorithm then reduces to the generation at each time step of an uniform random value $\chi\in (0,1)$, and its comparison with $p^{(1)}$. If $\chi\leq p^{(1)}$ one particle enters by the boundary, and otherwise no particle enters.

\subsubsection{Position of entering particles}
\label{sec-positions}

Once the number of particles entering into the system during the temporal step is decided, their initial position has to be determined. 
They cannot be placed at exactly the boundary, since they would leave the system by diffusion almost immediately.
Basically the initial position of new particles follows the probability (\ref{eq-rhoin}), once normalized by Eq.~(\ref{eq-nmean}). That is, the new positions of the entering particles have the probability density $p(x)$ given by
\begin{equation}
 \label{eq-p-in}
 p(x) = \frac{1}{2\,q(a)\sqrt{D\Delta t}} \erfc\left(\frac{x-f\Delta t}{\sqrt{4 D \Delta t}}\right).
\end{equation}

In order to sample the probability density, the standard method involves the inversion of the associated distribution function \cite{press07}. This distribution function is obtained by integration of the probability. The result for $p(x)$ is
\begin{equation}
 \label{eq-distrib}
 F(x)=\int_0^x p(x)\;dx = 1- \frac{q\left(\frac{x-f\Delta t}{\sqrt{4 D \Delta t}}\right)}{q(a)}.
\end{equation}
Then, for each new particle entering into the system one should obtain a uniform random number $\chi\in (0,1)$, and the desired initial position  is given by
\begin{equation}
\label{eq-sampling}
 x=F^{-1}(\chi).
\end{equation}

The actual inversion of $F(x)$ can be done numerically, for instance by saving in a table a series of values of $F(x)$ at the beginning of the simulation, and performing afterwards interpolations for the desired values of $F(x)$.
For large $x$ (i.e. $F(x)$ very close to 1) one can use the approximation 
\begin{equation}
%x\simeq f\Delta t+ \sqrt{- 4 \gamma k_BT \Delta t\ln(2\sqrt{\pi}q(a)(1-\chi)}
x\simeq f\Delta t+ \left({- 4 \Delta t \frac{k_BT}{\gamma}\ln(2\sqrt{\pi}q(a)(1-\chi))}\right)^{1/2}.
\end{equation}

% $y\simeq \sqrt{-\ln(2\sqrt{\pi}q(a)(1-F)}$ with the variable $y$ given by
% \begin{equation}
%  y = \frac{x-f\Delta t}{\sqrt{4 \gamma k_BT \Delta t}}.
% \label{y-change}
% \end{equation}

In our simulations we have employed Newton's method by using an approximate guess and by using the fact that the derivative of $F(x)$ is the function $p(x)$. In this way we can obtain the solucion of Eq.~(\ref{eq-sampling}) by the iteration
\begin{equation}
 y_{i+1}=\frac{1}{\erfc{y_i}}
 \left(\frac{1}{\sqrt{\pi}}e^{-y_i^2}-(1-\chi)q(a)\right)
\end{equation}
until the desired convergence is achieved. The initial position of the new particle is then obtained as
\begin{equation}
 %x=f\Delta t+ \sqrt{ 4 \gamma k_BT \Delta t} \,y
 x=f\Delta t+ \left(4 \Delta t \frac{k_BT}{\gamma}\right)^{1/2} \,y.
\end{equation}

\section{Simulation results and discussion}
\label{sec-results}

\subsection{Checking the algorithm}

\begin{figure}
\centering{\includegraphics[width=0.95\columnwidth]{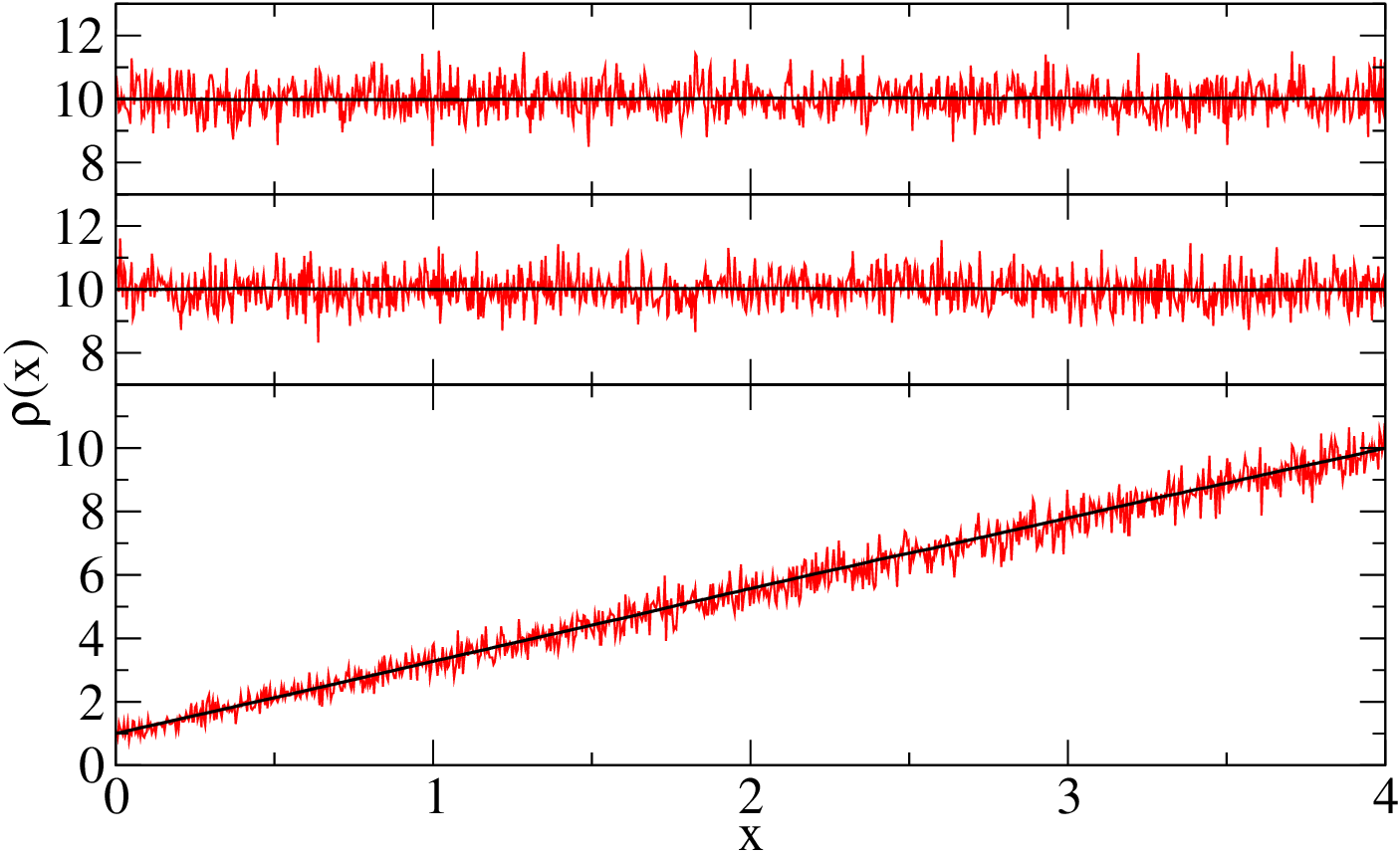}}
\caption{Particle concentration vs. position in the steady state for different values of the external potential $\phi$ and boundary conditions $\rho_1$, $\rho_2$. Red (gray) lines: average over 10000 independent realizations of the system; black lines: average on a single realization over a temporal window $T_\text{meas}=10^6$. Top: $q\phi=0$, $\rho_1=\rho_2=10$; middle: $q\phi=8 k_B T$, $\rho_1=\rho_2=10$; bottom:  $q\phi=0$, $\rho_1= 1$, $\rho_2=10$.
}
\label{fig-x-comp}
%\vskip5mm
\end{figure}
% 
% \begin{figure}
% \centering{\includegraphics[width=0.8\columnwidth]{10-10-phi-200.eps}}
% \caption{Particle concentration vs.~position in simulations with $q\phi=200$ and boundary conditions $\rho_1=\rho_2=10$. Thin lines: averages over temporal windows $T_\text{meas}=10^3$; thick red line: average over a temporal window $T_\text{meas}=10^4$; thick black line: average over a total measurement time $T_\text{meas}=10^6$.
% }
% \label{fig-10-10-200}
% %\vskip5mm
% \end{figure}

We have performed simulations of Brownian particles following the dynamics of Eq.~(\ref{eq-langevin}) in a channel of length $L=4$. The chosen temperature has been $k_BT=25$ and the friction parameter $\gamma=1000$. The temporal step has been taken as $\Delta t = 10^{-4}$. 
In a system of units based on length 1 nm, time 0.1 $\mu$s and energy 1 meV these values enter into the appropriate order of magnitude for ionic molecular channels. As a check of the algorithm we study the particle concentration along the channel and the resulting particle flux in different cases. Particle densities have been calculated by using $n=1000$ spatial bins of width $\Delta x=L/n$, and by counting the number of particles inside each bin at each step. 

In a first run of tests we have considered a constant deterministic drift $f(x)=f$, whose value has been varied. For convenience we write $f=-q\phi/\gamma L$, where $q\phi$ is the potential energy difference between both ends of the channel. Then $q$ and $\phi$ could be understood as the electric charge of the particle and the external electric voltage respectively. 
For this situation we have considered different values of concentration as boundary conditions at both ends of the channel.

We first simulate cases with the same value of concentration at both ends of the system,  $\rho_1=\rho_2=10$. In this case the steady particle distribution is a constant, equaling the value at the boundaries, for any value of the drift. Measured densities, however, are expected to present large fluctuations (since the particles are non-interacting, the number of them in each bin follows a Poisson distribution, from which it can be seen that values of local density should have a dispersion of order $\sigma_{\rho(\Delta x)} = \sqrt{\rho/\Delta x}$). It is hence most convenient averaging either by running many independent realizations of the system (reducing dispersion by $1/\sqrt{N}$) or by performing time averages in the steady state. This is shown in Fig.~\ref{fig-x-comp}. At the top we present an equilibrium situation with no external field applied (i.e. $q\phi=0$). The red (gray in the printed version) line is the average of 10000 independent realizations, taken once reached the steady state. The result presents the correct steady solution, with a dispersion of values $\sigma_i=0.512$ very close to the expected fluctuations $\sigma_{\rho(\Delta x)}/\sqrt{N} = 0.5$.
The black line is a long temporal average of a single realization in the steady state during a window $T_\text{meas}=10^6$. Here dispersion is strongly reduced, and it can be seen that results converge perfectly to the expected solution $\rho(x)= 10$. 

By imposing an external field ($q\phi=8 k_B T$) and maintaining the same boundary conditions we get the results shown in the middle of Fig.~\ref{fig-x-comp}. We also show averages over 10000 independent realizations [red (gray) line] and a long temporal average during a window $T_\text{meas}=10^6$ (black line).  We are no longer in an equilibrium situation, but the results agree as well with the steady state solution $\rho=\text{constant}$. The dispersion of values $\sigma_i=0.507$ is also very close to the expected value $0.5$.

This agreement is remarkable since the concentration value at the boundaries is imposed in the simulation by the mean rate at which new particles enter into the system only, which is calculated by using the boundary condition and the local value of the drift, but not using the expected flux or any other aspect of the solution of the problem. The dynamics of the particles (and their eventual exit from the system) is given exclusively by the trajectories of the Langevin equation (\ref{eq-langevin}), without using the value of the prescribed concentration or any other data.
It is also worth noting that there no boundary layer is visible near the ends, as could occur in some other approaches.

Note that by comparing the resulting particle densities in the cases without and with external field (top and middle plots in Fig.~\ref{fig-x-comp}), they are completely indistinguishable. They both present fluctuations that are uncorrelated in space, and of the same magnitude. Their local mean values are given by the solution of the Fokker-Planck equation (\ref{eq-fokker}), and for this equation the presence of a constant drift is equivalent to the switching to a moving reference frame. Therefore in this situation with homogeneous mean density the invariance of the results when changing the field was expected.

% \begin{figure}
% \centering{\includegraphics[width=0.8\columnwidth]{phi0-1-10.eps}}
% \caption{Particle concentration vs.~position in simulations with $q\phi=0$ and boundary conditions $\rho_1= 1$, $\rho_2=10$. Thin lines: averages over temporal windows $T_\text{meas}=10^3$; thick red line: average over a temporal window $T_\text{meas}=10^4$; thick black line: average over a total measurement time $T_\text{meas}=10^6$.
% }
% \label{fig-1-10-0}
% %\vskip5mm
% \end{figure}

We also test the steady solution reached by the system between boundaries with very different concentration values. For $q\phi= 0$ and $\rho_1= 1$, $\rho_2=10$ simulation results for local concentration are presented with the same statistics as the other cases at the bottom of Fig.~\ref{fig-x-comp}. In this case the steady solution between both values has a constant slope. Again results converge to the theoretical solution by means of a long temporal averaging. It is observed that fluctuations are larger in the higher concentration side, as expected, and reduced at the other boundary.

\begin{figure}
\centering{\includegraphics[width=0.95\columnwidth]{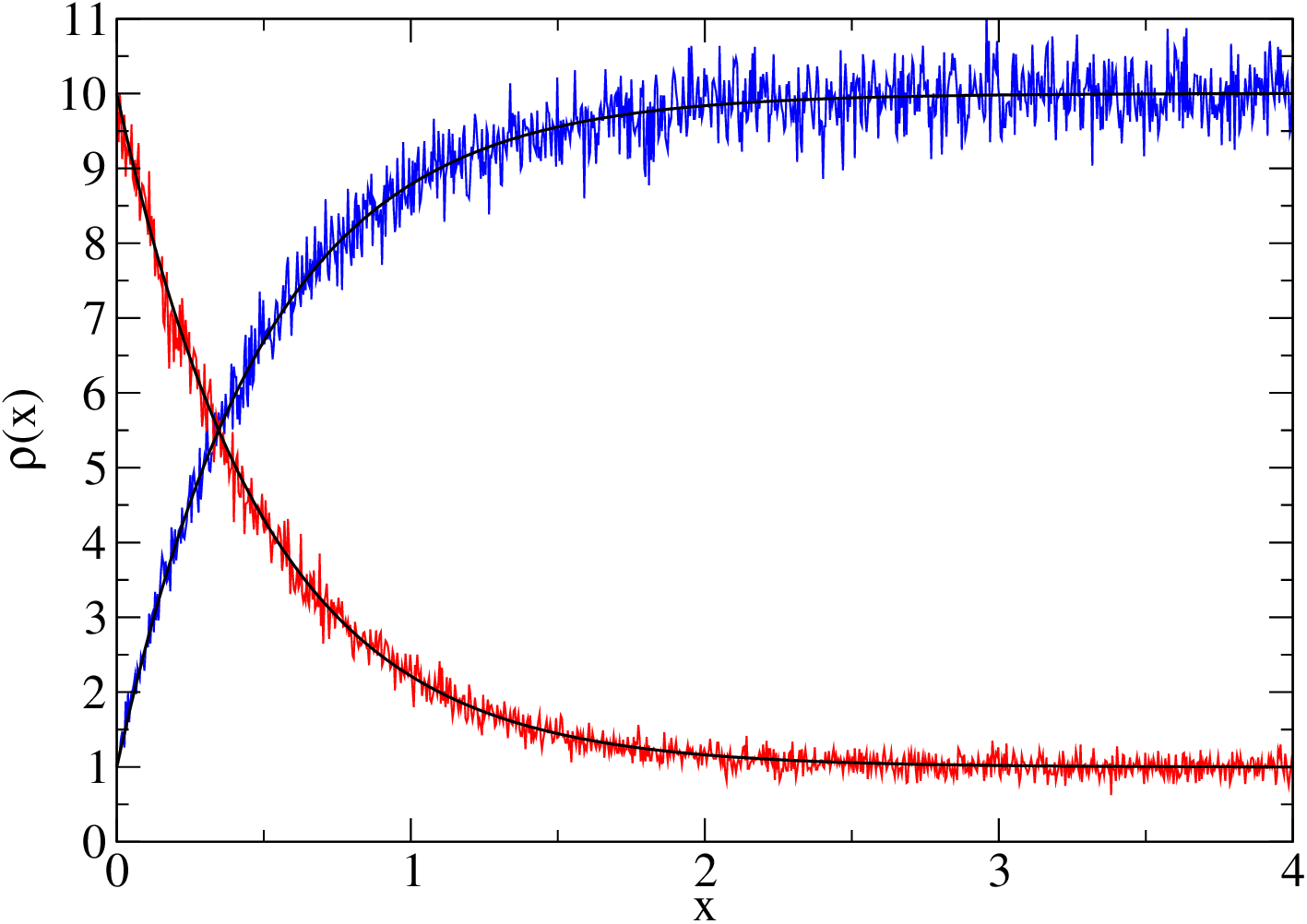}}
\caption{Particle concentration vs.~position in simulations with an constant external field $q\phi=8 k_B T$. Colored (gray) lines are simulation results averaged for $10000$ independent realizations; black lines are theoretical predictions from Eq.~(\ref{eq-steady}). Blue (dark gray) line: $\rho_1= 1$, $\rho_2=10$; red (light gray) line:  $\rho_1= 10$, $\rho_2=1$. 
}
\label{fig-steady-comp}
%\vskip5mm
\end{figure}

Simulation results and theoretical predictions are compared in Fig.~\ref{fig-steady-comp} for a large external field $q\phi =8 k_B T$ and boundary conditions with very different concentration values. Simulation results (fluctuating colored lines, gray in the printed version) are averages for 10000 realizations. Black lines correspond to the theoretical prediction of Eq.~(\ref{eq-steady}). In one case (blue, dark gray, line) both external field and concentration gradient drive the flow in the same direction, towards the left side, whereas in the other case both mechanisms act towards oposite directions.
As can be seen in both cases the agreement is very good. Long temporal averages for both cases have also been obtained (not shown), the results being virtually indistinguishable from the theoretical lines.

\begin{figure}
\centering{\includegraphics[width=0.95\columnwidth]{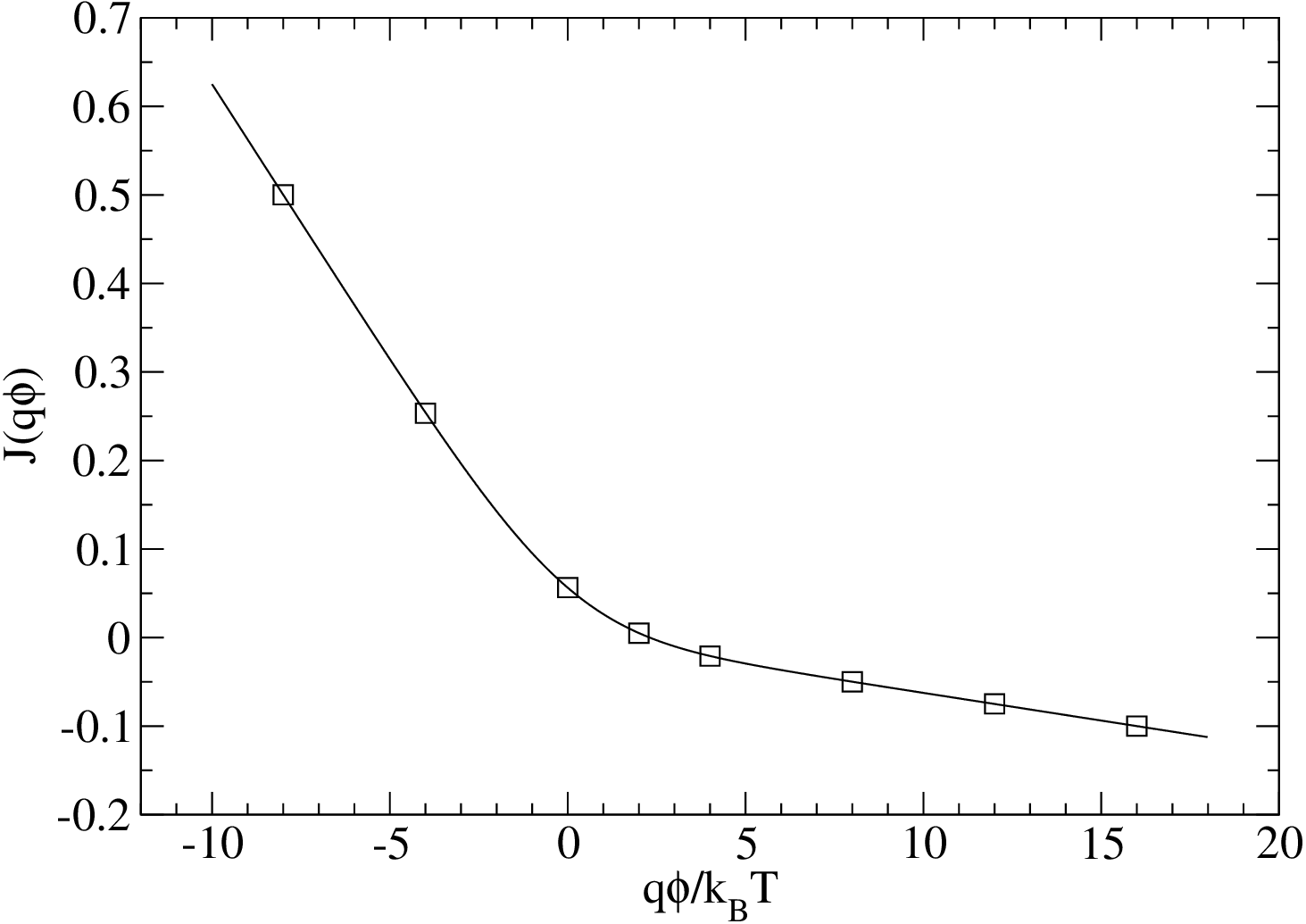}}
\caption{Particle flux (particles crossing the channel per unit time, averaged for a total time  $T_\text{meas}=10^6$) vs. potential energy, with boundary conditions $\rho_1=10$ and $\rho_2=1$. Line: theoretical prediction of Eq.~(\ref{eq-current}); symbols: simulation results. Observed differences in $J$ values between theory and simulations were in all cases smaller than $10^{-3}$.
}
\label{fig-j}
%\vskip5mm
\end{figure}

Next we have tested the prediction for the particle flux, Eq.~(\ref{eq-current}). For this test we have considered the situation with very different concentration values at both ends of the channel, $\rho_1=10$ and $\rho_2=1$, and performed simulations for different values of the drift, in order to span very different regimes. We have averaged the results over a time $T_\text{meas}=10^6$.
In Fig.~\ref{fig-j} we show theoretical predictions and simulation results for particle flux through the channel
We see that the agreement is virtually perfect in all regimes: namely for flux controlled by the concentration of a single end (i.e. very large voltages, either positive or negative) in which current is proportional to voltage, and for the crossover regime in which the flux is controlled by the concentration of both ends. In fact, by analyzing the obtained values, the differences with theoretical results are of the order of magnitude of the expected statistical uncertainties (since the crossing of individual particles forms independent events, the total number of them in a temporal span $T_\text{meas}$ is a Poisson process, which implies that the variance in the measured mean flux should be $\sigma_J^2=J/T_\text{meas}$).

\begin{figure}
\centering{\includegraphics[width=0.95\columnwidth]{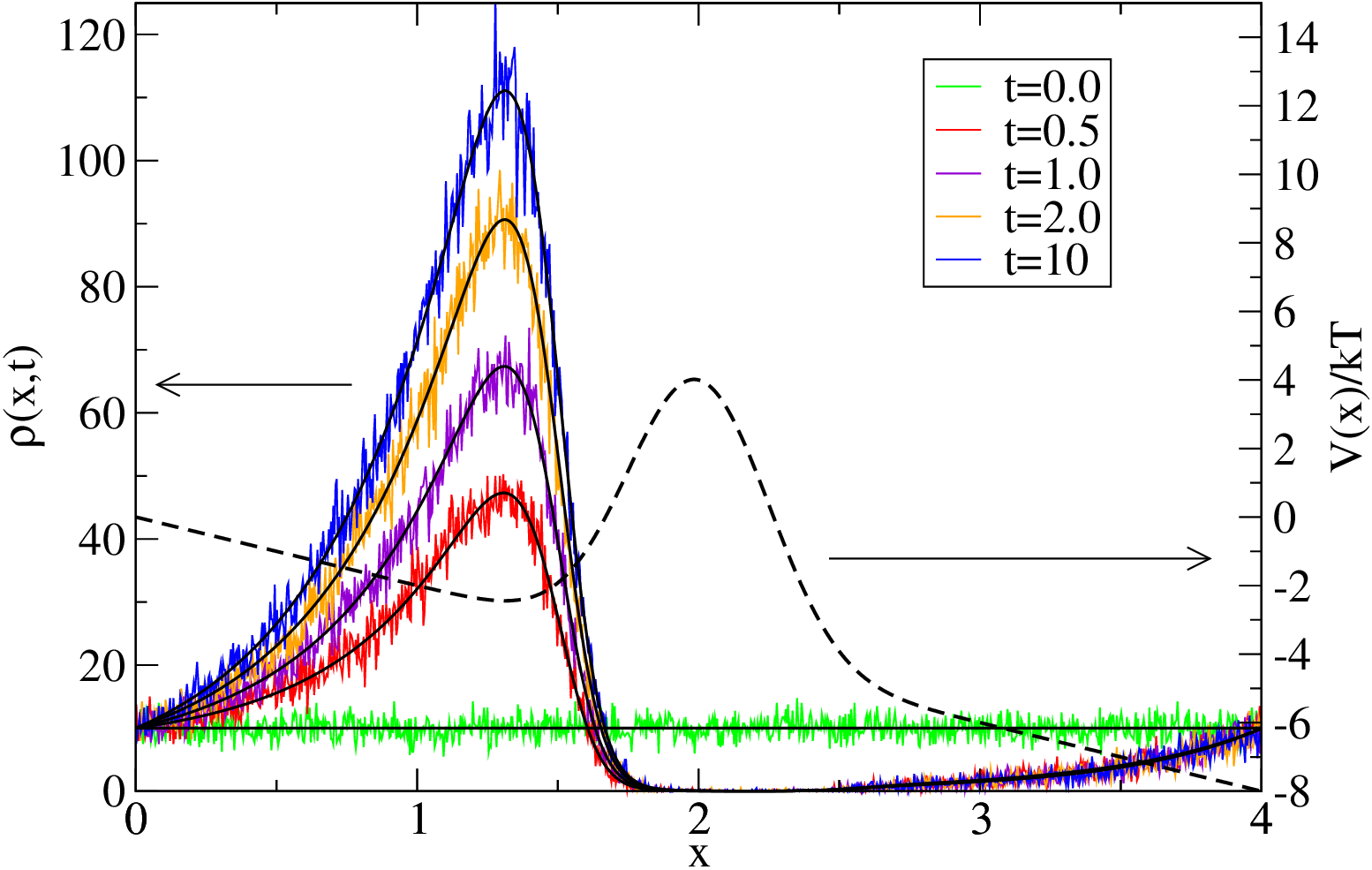}}
\caption{Particle concentration vs.~position during the transient ($t=0$, 0.5, 1, 2) and in the steady state ($t=10$). Simulations are performed with an external potential $q\phi=-8 k_B T$ and %the presence of 
a potential barrier as specified in Eq.~(\ref{eq-barrier}), also represented. %The parameters of the barrier are $V_b = 8 K_B T$, $l=L/16$, and $x_b=L/2$. 
%The employed initial conditions were prepared by taking configurations in the steady state of the same case but without barrier. 
Colored (gray) lines: simulation results averaged for $10000$ independent realizations; black lines: numerical integration of Eq.~(\ref{eq-fokker}); dashed line: applied potential acting on the particles.
}
\label{fig-barrier}
%\vskip5mm
\end{figure}

We finally simulate the case of a channel with a potential barrier in the middle. This could be representative of an ionic channel with a more complex potential landscape, or with the presence of a gate. Following the shape of the gates used in Refs.~\cite{ramirez1,ramirez2,ramirez-periodic} we add a Gaussian barrier to the constant drift, with which the total potential reads
\begin{equation}
 V(x)=q\phi(x-x_1)/L + V_b \exp -\frac{(x-x_b)^2}{2 l^2},
 \label{eq-barrier}
\end{equation}
where $V_b$ is the height, $l$ is the width, and $x_b$ is the center position of the barrier. We have taken the values $V_b = 8 K_B T$, $l=L/16$, $x_b=L/2$, and the potential energy difference $q\phi=-8 k_B T$. In this case the constant force term pushes the particles towards the right side, but they are trapped due to the barrier and the resulting flow is reduced. As a result one observes a slow transient during which a large number of particles are being accumulated until the system reaches the steady state. We have followed this transient and compared it with the direct numerical integration of the Fokker-Planck equation (\ref{eq-fokker}). Results are presented in Fig.\ref{fig-barrier}. We see in this figure that the agreement is very good during the entire transient. It is remarkable that the rate at which particles are accumulating depends critically on the balance of entering and exiting particles, and hence on the boundary conditions.

\subsection{Boundary conditions revisited}

The proposed algorithm for the fixed concentration boundary conditions belongs to the class of methods that assign probabilistic rules for the appearance of new particles at the boundaries \cite{eisenberg95,nadler05,singer05}. An important difference of this type of algorithms from others using buffer zones or particle reservoirs is that in the latter methods the dynamics at the boundary outside the system is simulated with buffer particles, which implies that some assumptions on this dynamics (for instance the size of the buffer zone, or the criteria for maintaining constant concentration) have to be established. The validity of these assumptions should be discussed in terms of the real physics of the system at hand. 

In fact, also in the present algorithm (and in other algorithms assuming the probabilistic appearance of particles at the boundary) an important assumption has been implicitly assumed:
the probability of new particles entering into the system does not depend on past exit events, and as a consequence the boundary loses memory of any exiting particle instantaneously. This is equivalent to saying that the particle reservoirs (which are not explicitly simulated) are assumed to be mixed at each time step. An immediate effect is an implicit loss of temporal correlations near the boundaries.
We will illustrate the consequences of this assumption by returning to the simple simulations of the channel without a barrier and with the same values of prescribed concentrations at both ends of the channel.

\begin{figure}
\centering{\includegraphics[width=0.95\columnwidth]{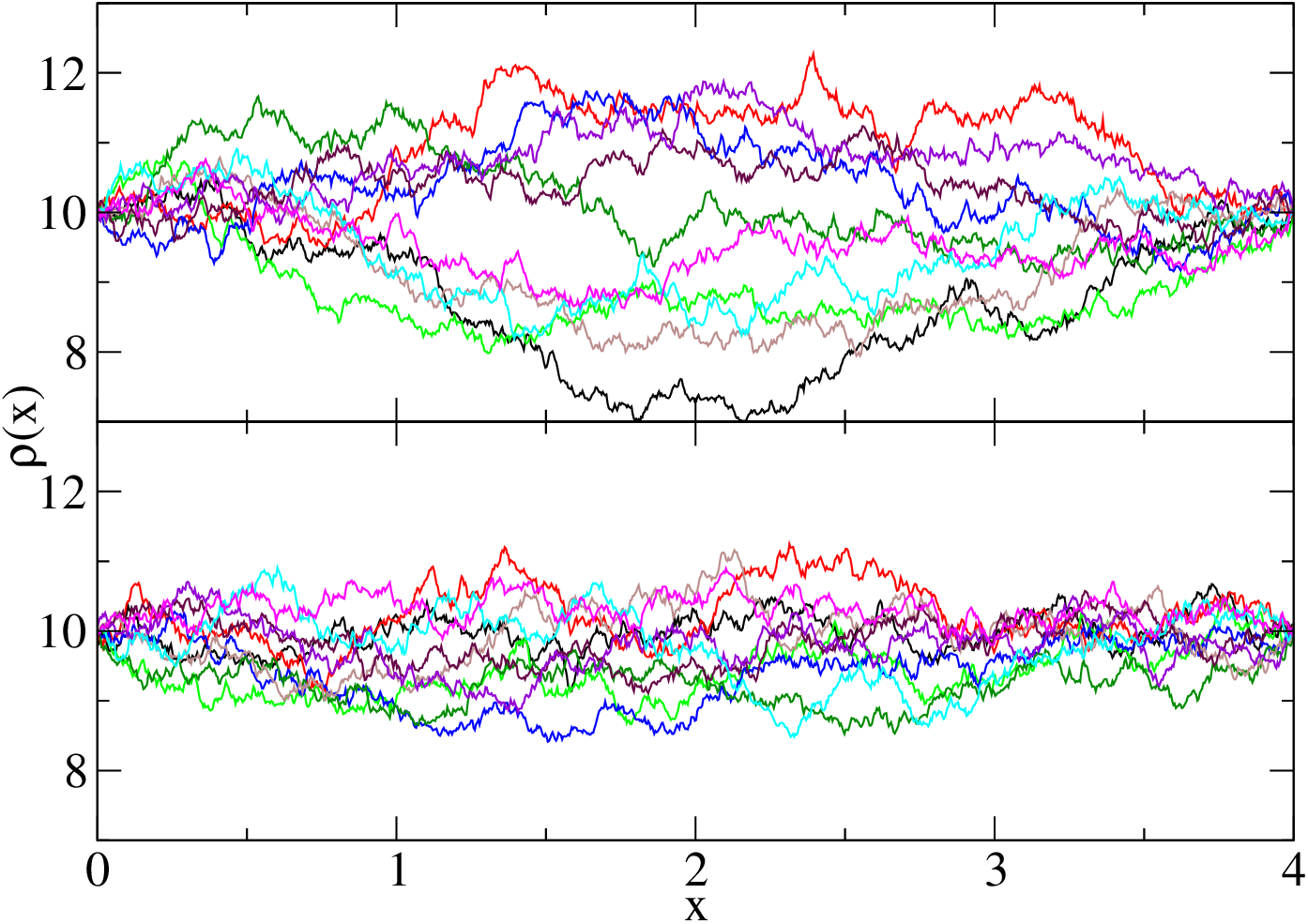}}
\caption{Particle concentration vs.~position for temporal averages $T_\text{meas}=10^3$ in the steady state. Top: $q\phi=0$; bottom $q\phi=8 k_B T$.
}
\label{fig-temp}
%\vskip5mm
\end{figure}

We show again in Fig.~\ref{fig-temp} (top) results for the equilibrium situation with no external field applied ($q\phi=0$) as in Fig.~\ref{fig-x-comp} (top). Here we show several temporal averages calculated in shorter windows $T_\text{meas}=10^3$. We clearly see here that when performing temporal averages the appearance of the fluctuations is radically different from when performing averages over independent realizations. Namely the concentration values in different positions are correlated when they are calculated by temporal averages, due to the contribution of the same particles moving diffusively across the system, which results in the shape of the fluctuations appearing in the figure. Short length fluctuations are short lived (their associated diffusion times 
have smaller diffusion times $\sim\lambda^2/D$) and they are averaged out.
This is so because the effective number of independent observations of a statistical quantity with correlation time $T_\text{corr}$ in a temporal window $T_\text{meas}$ is of the order of $T_\text{meas}/2T_\text{corr}$, and the resulting dispersion is reduced as the square root of this effective number (see, for instance, Ref. \cite{Heermann90}).
However long length fluctuations are slower and do not disappear in not very long temporal averages. This is manifested in the large dispersion of results that can be seen at the center of the channel. 

This dispersion of results is strongly reduced at the boundaries, as can also be seen in Fig.~\ref{fig-temp} (top). Since the loss of temporal correlations is produced at the boundaries, fluctuations there will have shorter correlation times than in the middle of the system. As a result statistical dispersion after temporal averaging will also be shorter near the boundaries, and the averaged concentration values will better converge to the prescribed values of boundary conditions there.
When performing even shorter temporal averages (not shown) dispersions are seen to be larger in the middle of the system but they are also reduced at both ends, in such a way that the prescribed concentration values are also verified.

In Fig.~\ref{fig-temp} (bottom) we present the case with the same concentration at the boundaries, but with an applied field $q\phi= 8 k_B T$. %Similarly to the equilibrium case
Differently from what is observed in Fig.~\ref{fig-x-comp}, temporal averages are no longer invariant when changing the field.
Dispersion at the ends are also reduced and the boundary conditions are well verified, but dispersion in the middle of the channel is smaller than in the equilibrium case. 
%In this case the time correlation of the fluctuations is limited by the ballistic time $L/f$.
This is due to the deterministic drift given by the external field. Particles in average remain in the system a ballistic time of the order of $L/q\phi$, smaller than the diffusion time, and therefore there is a shorter time to develop density fluctuations. The lifetimes of the largest fluctuations are shorter than in the equilibrium case and, as a consequence, the same simulation times permit a better average and the observed dispersion is smaller.

These observations constitute an indication that in Langevin simulations of diffusing particles, constant concentration boundary conditions imply more assumptions that the mere specification of the mean concentration values at the boundaries. 
To illustrate this point let us consider the case of solving the simplest case of a small piece or segment of length $L$ of a very large (or infinite) channel with a mean density $\rho$ of free particles in equilibrium. At the level of the Fokker-Planck equation for the mean density the appropriate boundary condition will be the fixing of the value of the concentration at both ends of the segment of length $L$,  i.e. the same as what has been considered here. But at the level of the Langevin dynamics of the particles, one should expect in this case the presence of density fluctuations of large scale (much larger than the piece size $L$), which will decay very slowly, at times of the order of its diffusion time. Then, in a finite time averaging the piece of length $L$ should almost always be in a large fluctuation, and the concentration values of the boundary conditions would almost never be observed. 

On the contrary, the way in which we have implemented the boundary conditions in this work (corresponding to the same Fokker-Planck equation and with the same solution for the mean density) does not correspond to a piece of a such much larger homogeneous system, but instead to a system of length $L$ between two boundaries separating the system from reservoirs that are perfectly mixed. It is precisely the loss of correlations at the boundaries which permits the prescribed concentration values to be observable in temporal averages.

\section{Concluding remarks}

We have derived an algorithm for the integration of first order additive Langevin equations in the presence of fixed concentration boundary conditions. 
% This could be understood as the presence of particle reservoirs, which are always homogeneously mixed with the prescribed concentrations, in contact with the boundary.
The method consists of letting disappear any particle that exit from the system through the boundary, and the appearance of new particles near the boundaries according to the prescribed condition. For this we have calculated the appearance probability and the distribution of initial positions for the new particles. This method implies that particle memory is lost when crossing the boundary, which is equivalent to resetting the positions of the reservoir particles at each time step. This permits us to eliminate fluctuations larger than the system size, which would hinder the verifying of the boundary condition in finite simulations.

In situations in which there is not a perfect mixing in the reservoirs, the method would also be appropriate for modeling separately the microscopic dynamics of the individual particles in the system (i.e. the channel) and the concentration dynamics outside the system (for instance by means of more mesoscopic or hydrodynamic models). The resolution of the latter dynamics would then be used as the boundary condition for the former. The only requirement in this case would be that the resulting flux across system had a negligible effect over the reservoirs. This is the important case of a very thin channel joining two large cavities.

This algorithm for the boundary conditions is closely related to what employed in other works \cite{eisenberg95,nadler05,singer05}, where the simulation of the particle reservoirs was also avoided by considering the stochastic appearance of new particles through the boundary. In particular in Ref. \cite{singer05} it was already remarked that some prescription was necessary for the initial position of the new particles in order to avoid the appearance of spurious depletion boundary layers at the boundary, and a specific distribution of new positions was derived provided the flux was known. This fact limited the procedure to analytically resolvable problems or to resorting to a shooting method to adjust a simulation parameter.
On the contrary for our algorithm we have directly calculated both appearance and new position probabilities, depending only on the desired boundary condition and on the local drift at the boundary. As a result we obtain in all cases the correct results without the use of any adjustable parameter.

We have checked simulation results in conditions appropriate for modeling molecular ionic channels at the cellular membrane, that is for the Brownian dynamics in the overdamped limit of ions with a constant deterministic drift corresponding to the value of the membrane potential. Results agree perfectly with theoretical predictions, both for (steady and transient) density distribution and for resulting mean flux. Moreover the magnitude of fluctuations agree with the expected values.

As already pointed out, the method has no adjustable parameter, and does not involve the explicit simulation of particles in the reservoirs or the consideration of velocity variables that should be irrelevant in the low Reynolds limit. The results do not show any trace of residual boundary layers. It can be applied to any model formulated in terms of trajectories of Brownian particles following first order additive Langevin equations of the type of Eq.~(\ref{eq-langevin}). Therefore it is specially appropriate in Langevin approaches for the study of molecular channels and the modeling of the interaction between gates and individual ions and the action of the membrane potential, as used in Refs. \cite{ramirez1,ramirez2,ramirez-periodic}. 

\section*{Acknowledgments}

This work was performed in the context of an extensive collaboration with Prof. J.M. Sancho in the area of molecular channel dynamics, and many and very fruitful discussions with him are gratefully acknowledged.
This work was supported by the Ministerio de Economia y Competividad (Spain) and FEDER (European Union), under
project FIS2015-66503-C3-2-P.

This paper was published as Phys. Rev. E 98, 013302 (2018) with missprints in Eqs. (17) and (19). These missprints have been corrected here. We would like to thank Josep Bataller, Guillermo Villanueva, and Ce Xu Zheng for reporting numerical inconsistencies when using the published equations. 
%This work was supported by the Spanish DGICYT Projects No. FIS2012-37655 
%and by the Generalitat de Catalunya Projects 2009SGR14 and 2014SGR878. 
 \vspace*{6pt}

\bibliography{references}

\end{document}